\documentclass[10pt,aps,prl,twocolumn,superscriptaddress]{revtex4-2}
\usepackage{graphicx}
\usepackage[colorlinks, linkcolor=blue]{hyperref}
\hypersetup{colorlinks,allcolors=blue}
\usepackage{physics}

\usepackage[capitalise]{cleveref}

\newcommand{\bcen}{\begin{center}}
\newcommand{\ecen}{\end{center}}
\newcommand{\btab}{\begin{tabular}}
\newcommand{\etab}{\end{tabular}}
\newcommand{\bdes}{\begin{description}}
\newcommand{\edes}{\end{description}}

\newcommand{\beq}{\begin{equation}}
\newcommand{\eeq}{\end{equation}}
\newcommand{\bea}{\begin{eqnarray}}
\newcommand{\eea}{\end{eqnarray}}

\newcommand{\bary}{\begin{array}}
\newcommand{\eary}{\end{array}}
\newcommand{\benum}{\begin{enumerate}}
\newcommand{\eenum}{\end{enumerate}}
\newcommand{\bitem}{\begin{itemize}}
\newcommand{\eitem}{\end{itemize}}

\newcommand{\Fig}[1]{Fig.~\ref{#1}}

\makeatletter

\newcommand{\Rmnum}[1]{\expandafter\@slowromancap\romannumeral #1@}
\makeatother

\begin{document}

\title{Topologically switchable transport in a bundled cable of wires}


\author{Nirnoy Basak}
\email{nirnoyb@iitk.ac.in}
\affiliation{Department of Physics, Indian Institute of Technology Kanpur, Kalyanpur, Uttar Pradesh 208016, India}

\author{Ritajit Kundu}
\email{ritajit@iitk.ac.in}
\affiliation{Department of Physics, Indian Institute of Technology Kanpur, Kalyanpur, Uttar Pradesh 208016, India}
\affiliation{Institute of Mathematical Sciences, CIT Campus, Chennai 600113, India}
\affiliation{Homi Bhabha National Institute, Training School Complex, Anushaktinagar, Mumbai 400094, India}

\author{Basudeb Mondal}
\affiliation{Department of Physics, Indian Institute of Science, Bangalore 560012, India}

\author{Adhip Agarwala}
\email{adhip@iitk.ac.in}
\affiliation{Department of Physics, Indian Institute of Technology Kanpur, Kalyanpur, Uttar Pradesh 208016, India}

\begin{abstract}
Advances in the next generation of mesoscopic electronics require an understanding of topological phases in inhomogeneous media and the principles that govern them. Motivated by the nature of motifs available in printable conducting inks, we introduce and study quantum transport in a minimal model that describes a bundle of one-dimensional metallic wires that are randomly interconnected by semiconducting chains. Each of these interconnects is represented by a Su-Schrieffer-Heeger chain, which can reside in either a trivial or a topological phase. Using a tight-binding approach, we show that such a system can transit from an insulating phase to a robust metallic phase as the interconnects undergo a transition from a trivial to a topological phase. In the latter, despite the random interconnectedness, the metal evades Anderson localization and exhibits a ballistic conductance that scales linearly with the number of wires. We show that this behavior originates from hopping renormalization in the wire network.  The zero-energy modes of the topological interconnects act as effective random dimers,  giving rise to an energy-dependent localization length that diverges as $\sim 1/E^2$. Our work establishes that random networks provide a yet-unexplored platform to host intriguing phases of topological quantum matter.     

\end{abstract}

\maketitle

{\it Introduction:} Topological phases of matter and their classification have ushered in a new paradigm in the understanding of quantum phases of matter where non-trivial band topology leads to quantized transport signatures \cite{Altland_PRB_1997, Kane_Z2_2005, BHZ_Sci_2006, Konig2007QSH, Kitaev_AIP_2009, Moore_Nat_2010, Qi_RMP_2011, Ludwig_PS_2015}. Remarkable phenomena in a range of systems including quantum Hall \cite{Hatsugai_IOP_1997,Tsui_1982, Thouless_PRL_1982,  Kane_PRL_2005}, topological insulators \cite{ Fu_PRL_2007,  Hasan_RMP_2010, Qi_RMP_2011, Shen_Book_2013,Bernevig_Book_2013} and twisted-layered materials \cite{Gail_PRB_2011, Song_PRL_2019, Nuckolls_Nat_2020, Wu_Nat_2021} directly result through such band topology. However, in recent years, understanding of topological phases have been extended to amorphous networks \cite{Agarwala_PRL_2017,Mitchell_NatPh_2018, costa_amorphusti_2019, zhou_amorphusTI_2020, Marsal_PNAS_2020, Zhang_amorphustopology_science_2023,   Corbae_Nat_Mat_2023, Corbae_EDPSc_2023}, disordered materials \cite{ Jian_PRL_2009, Jiang_PRB_2009, Groth_PRL_2009, Guo_PRL_2010,Chaou_PRB_2025}, trees \cite{ManojPRB2023, singh2024arboreal}, photonic \cite{Khanikaev_Nat_2017, Noh_Nat_2018, Peterson_Nat_2018,  Ozawa_RMP_2019, Fu_OQE_2020} and synthetic systems \cite{Seo_PRL_2012, Liu_PRL_2013, Galitski_Nat_2013, Cooper_RMP_2019, Ni_ChemR_2023}. Possibilities of engineering defects \cite{Teo_PRB_2010, Pham_PRB_2019, Chen_Nat_2025} and disorder \cite{ Cardano_Nat_2017, Xu_NatCom_2017, Ling_PRB_2024} to tune material properties of such topological systems towards designing tunable functionalities hold immense promise. An interesting platform in this respect has been those of printable polymers, where the microscopic electronic motifs comprise randomly strewn conducting wires \cite{Cao_2008}. These have been of immense interest to the next generation of flexible electronics and biosynthetic applications \cite{Wang_2010,Sun_2011,Duan_2013,Lee_2014,Manoranjan_2025}. However, theoretical understanding of quantum transport through such systems, or even their microscopic modeling, is rudimentary and in a premature stage. 

\begin{figure}[h!]
\centering
\includegraphics[width=0.85\linewidth]{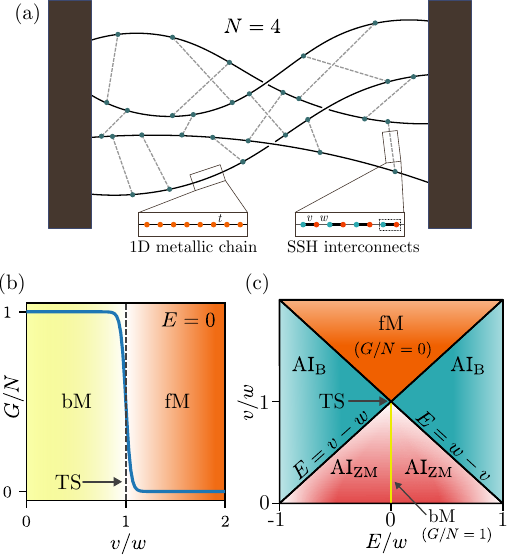}
\caption{\textbf{Model and Phase Diagram:}
(a) Schematic of $N$ one-dimensional metallic wires (solid lines) randomly connected by SSH interconnects (dashed lines) between two rectangular contacts. (b) At $E=0$, the two-terminal conductance per wire ($G/N$) is unity for $v/w<1$ (topological) and vanishes for $v/w>1$ (trivial). We refer to this as \emph{topological switching} (TS).
For $v < w$, the $G/N=1$ regime is the ballistic metal (bM).
When $v>w$, the interconnects effectively break the wires, resulting in \emph{fractured metal} (fM). 
(c) Schematic phase diagram in the $(E/w, v/w)$ plane. When $E\neq0,$ the system evolves into an Anderson insulator (AI) induced by either SSH zero modes ($\mathrm{AI_{ZM}}$) or due to SSH bulk bands, denoted $\mathrm{AI_{B}}$ ($t=1$).} 
\label{fig:fig1}
\end{figure}

In this work, we model a bundle of one-dimensional wires that are randomly interconnected by a soup of topological interconnects. We pose, as the topological properties of interconnects are tuned, can one modulate the transport properties of the metallic wires? The system is illustrated in \Fig{fig:fig1}(a) where the metallic wires are characterized by one-dimensional tight-binding chains with the hopping scale $t=1$. The insulating interconnects are characterized by Su-Schrieffer-Heeger (SSH)\cite{Su_1979_PRL} chains with dimerized hoppings $v$ and $w$. We maintain the metallic chains at a chemical potential $E$, such that at half-filling, $E=0$. The central result of our work is summarized schematically in \Fig{fig:fig1}(b) and (c). 

We find that when these interconnects are in a trivial phase ($v > w$), the wires effectively fracture into metallic segments, resulting in a vanishing two-terminal conductance. We dub this phase as fractured metal ($\text{fM}$). On the other hand, when the interconnects are in the topological phase ($v<w$), this results in pristine metallic transport through the wires, which is dissipation-free. We dub this phase as ballistic metal (bM). Interestingly, even though one-dimensional wires are expected to generically Anderson localize in the presence of any infinitesimal disorder \cite{Anderson_PR_1958}, here - $N$ wires, although randomly interjected through topological junctions, do not result in any scattering, leading to a two-terminal ballistic conductance with $G/N=1$. We dub this phenomenon as `topological switching' (TS), where a topological transition of insulating interconnects between metallic wires induces a transition from an insulating fractured metal to a robust dissipation-free metal (see \Fig{fig:fig1}(b)). Furthermore, away from the half-filling ($E \neq 0$), the system generically localizes, but the scatterings are either dominated via the zero-modes of the SSH chains ($\text{AI}_\text{ZM}$) or via their bulk ($\text{AI}_\text{B}$). Next, we directly delve into the model.

 \begin{figure*}
    \centering
    \includegraphics{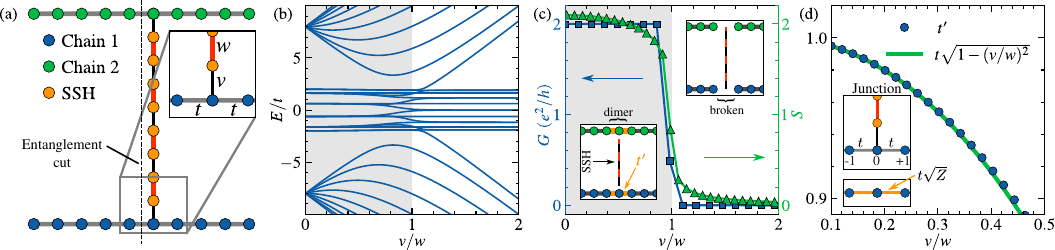}
\caption{
\textbf{Characterizing the metallic chain and SSH junction:}
(a) Junction geometry: two metallic chains connected through an SSH segment; the inset illustrates the chosen hopping amplitudes. The dashed line marks the entanglement cut. 
(b) Energy spectrum of the junction as a function of $v/w$, computed with periodic boundary conditions and parameters $L = 10$ and $L_{\mathrm{SSH}} = 20$. 
(c) Conductance and entanglement entropy as functions of $v/w$, using $L = 100$ and $L_{\mathrm{SSH}} = 80$. The entanglement entropy is evaluated at the cut shown in panel (a). 
The insets highlight the qualitative change in the Wannier tight-binding description: for $v/w < 1$, a dimer forms at the junction, whereas for $v/w > 1$ the junction site effectively disappears and the chains become fragmented. 
(d) Comparison between the dimer hopping $t'$ extracted from the Wannier tight-binding model and the analytical expression, using $L = 10$ and $L_{\mathrm{SSH}} = 10$. The inset shows the indexing convention of the junction and the effective renormalization of the hopping $t \to t \sqrt Z$. All calculations in this panel use $w/t = 8$ and $t = 1$. }
    \label{fig:fig2}
\end{figure*}

{\it Model:} This tight-binding Hamiltonian is given by (see \cref{fig:fig1}(a)):
    \begin{align}
        H = H_{\mathrm{W}} + H_{\mathrm{SSH}} + H_{\mathrm{C}}
        ,
        \label{eq:totalHam}
    \end{align}
    where,
    \begin{align}
        H_{\mathrm{W}} = &-\sum_{n=1}^{N}\sum_{i=1}^{L-1}tc^{\dag}_{n,i}c_{n,i+1} + \mathrm{H.c.},\nonumber\\
        H_{\mathrm{SSH}} = &-\sum_{m=1}^{N_{\mathrm{SSH}}}\Bigg(\sum_{j=1}^{\frac{L_{\mathrm{SSH}}}{2}-1}wb^{\dag}_{m,j}a_{m,j+1}\nonumber\\
        &\qquad\qquad+\sum_{j=2}^{\frac{L_{\mathrm{SSH}}}{2}-1}vb^{\dag}_{m,j}a_{m,j}\Bigg) +\mathrm{H.c.}\nonumber\\
        H_{\mathrm{C}} = &-v \sum_{m=1}^{N_{\mathrm{SSH}}}
        \left(c_{I(m)}^{\dag}b_{m,1} + 
        c_{J(m)}^{\dagger} a_{m,\frac{L_{\mathrm{SSH}}}{2}} \right)
        + \mathrm{H.c.}\nonumber
        \end{align}
\(H_{\mathrm W}\) is the Hamiltonian of \(N\) one-dimensional metallic wires, each of length \(L\).
The operator \(c_{n,i}\) annihilates a fermion on the \(i\)-th site of the \(n\)-th wire, with nearest-neighbor hopping amplitude \(t\). \(H_{\mathrm{SSH}}\) describes \(N_{\mathrm{SSH}}\) SSH interconnects, each of length \(L_{\mathrm{SSH}}-2\). The operators \(a_{m,j}\) and \(b_{m,j}\) annihilate fermions on the two sublattices of the \(j\)-th unit cell of the \(m\)-th SSH chain. $v$ and $w$ represent the inter- and intra-cell hopping amplitudes of the SSH interconnects, respectively. For each SSH chain, the first and last sites are coupled to a randomly chosen metallic wire at a random site, with a hopping amplitude of \(v\). These couplings are described by \(H_{\mathrm C}\), which couple the bottom and top ends of the SSH chain to the wires. \(I(m)\) specifies the wire and site connected to the top end of the \(m\)-th SSH chain, while \(J(m)\) specifies the corresponding wire and site for the bottom end, which are generated randomly.  
A total of \(N_{\mathrm{SSH}}\) SSH chains connect \(2N_{\mathrm{SSH}}\) randomly chosen sites on the metallic wires, leading to an average interconnect density \(\rho = 2N_{\mathrm{SSH}}/(N L)\).

The system possesses both chiral and time-reversal symmetry and thus belongs to the symmetry class BDI \cite{Altland_PRB_1997,Chiu_RMP_2016}. We focus primarily on half-filling, corresponding to \(E = 0\).
The metallic wires are characterized by a Fermi velocity \(v_F \sim 2t\), while the SSH interconnects have a spectral gap \(\Delta \sim |v - w|\) and an associated correlation length \(\xi_{\mathrm{SSH}} \sim 1/\Delta\).

{\it Single Junction:} 
To understand the properties of the full network, it is useful to first study a system of two metallic chains (Chain~1 and Chain~2) connected by a single SSH interconnect as shown in \Fig{fig:fig2}(a). The end sites of the SSH chains, which would host a zero-energy edge mode in the topological phase, are incorporated into the metallic wires. Periodic boundary conditions (PBCs) are imposed along the wires. The resulting single-particle spectrum of the two-wire junction is shown in \Fig{fig:fig2}(b). In the regime \(v \ll w\), where the SSH interconnect is deep in the topological and \(w \gg t\), the spectrum exhibits a clear separation of energy scales. Near \(|E| \lesssim 2t\), there are \(L\) states that are predominantly supported on the metallic wires. Surprisingly, the zero-energy edge mode of the SSH interconnect does not appear in this spectrum.  Instead, the states in this energy range resemble those of two independent chains of length $L$  with PBCs. At higher energies, \(|E| \gtrsim |v - w|\), there are \(L_{\mathrm{SSH}} - 2\) states that are predominantly supported on the bulk of the SSH interconnect.

The situation is drastically different when $v\gg w$, when the SSH chain is deep in the trivial limit. In this case the the spectrum at $\abs{E} \gtrsim \abs{v-w}$ has now $L_{\mathrm{SSH}}$ number of states, whereas the spectra in the $|E|\lesssim 2t$ range are similar to those of two independent chains with open boundary conditions with length $L-1$, despite the PBCs imposed on the wires. This indicates that the junction effectively fractures the chains. 
  
To confirm this physical picture, we compute the bipartite entanglement entropy \(S\), with the entanglement cut shown in \Fig{fig:fig2}(a). As \(v/w\) is tuned across the topological phase transition, \(S\) changes from $\sim 2$ to $zero$, indicating that each of the two wires becomes disconnected at the entanglement cut. This behavior is corroborated in the two-terminal conductance along the wires calculated using the non-equilibrium Green’s function formalism \cite{Datta_Book_1997}, which decreases from \(G = 2 e^2/h\) to zero [see \Fig{fig:fig2}(c)]. These results suggest that the SSH interconnect does not lead to any back-scattering in the topological regime (\(v < w\)), whereas it effectively fractures the metallic wires in the trivial regime (\(v > w\)), as illustrated schematically in the insets of \Fig{fig:fig2}(c).

{\it Wannierization:} 
To quantitatively understand the physics at the junction, we Wannierize and reconstruct the effective tight-binding Hamiltonian for the metallic wires within the energy window $\abs{E} \lesssim 2t$.  We first diagonalize the Hamiltonian, \(H = U D U^\dagger\), where \(U_{i n} = \braket{i}{\psi_n}\). Here \(\ket{i}\) denotes the orbital basis, \(\ket{\psi_n}\) are the eigenstates of \(H\), and \(D\) is a diagonal matrix with entries \(D_{n n} = \varepsilon_n\).

To construct Wannier functions, we introduce a complex representation of the orbital positions \((x_j,y_j)\), \(z_j = y_j \exp\!\left( 2\pi i x_j / L \right)\). The index \(j\) runs over all orbitals in the same order as used to construct the total Hamiltonian in ~\cref{eq:totalHam}. The coordinates satisfy \(x_j \in [0,L-1]\) and \(y_j \in [1,L_{\mathrm{SSH}}]\). This representation respects PBCs along the \(x\)-direction. We then project the position operator onto the subspace of eigenstates within the metallic energy window, \(X_{p q} = \sum_j z_j \braket{\psi_p}{j} \braket{j}{\psi_q}\), where \(p,q\) label eigenstates with energy \(|\varepsilon_p|, |\varepsilon_q| \lesssim 2t\). Diagonalizing the projected position operator, \(X = V \Lambda V^\dagger\), yields the maximally localized Wannier orbitals (MLWOs) \(\ket{w_i} = \sum_n V_{n i} \ket{\psi_n}\). Here \(V_{n i} = \braket{\psi_n}{w_i}\), and \(\Lambda\) is a diagonal matrix with entries \(\Lambda_{n n} = \lambda_n\). The complex numbers \(\lambda_n\) encode the Wannier-center positions, given by \(x_n = |\lambda_n|\) and \(y_n = L \arg(\lambda_n) / 2 \pi\). Finally, the Wannierized tight-binding Hamiltonian is \([\mathcal H_{\mathrm{w}}]_{m n} = \matrixel{w_m}{H}{w_n}\).

From the Wannier Hamiltonian, we find that in the topological regime \(v < w\), the effective hopping between the site connected to the SSH interconnect and the rest of the metallic wire is renormalized to a reduced value \(t'\) [see lower inset of \Fig{fig:fig2}(c)]. This renormalization can be understood analytically. In the topological phase of the SSH interconnect, a zero-energy edge mode forms with dominant weight on the site shared with the metallic wire, denoted \(\ket{0}\). The edge-state wavefunction takes the form \(c_e^\dagger \ket{\Omega} = \ket{e} \sim \sqrt{Z}\,\ket{0} + \ldots\), where the omitted terms decay exponentially into the bulk of the SSH interconnect. As a result, the hopping near the junction, \(-t (c_0^\dagger c_{+1} + c_0^\dagger c_{-1} + \text{H.c.})\), is renormalized to \(-t \sqrt{Z} (c_e^\dagger c_{+1} + c_e^\dagger c_{-1} + \text{H.c.})\), where \(c_e^\dagger\) creates the effective edge mode. A direct analytical calculation yields \(Z = 1 - (v/w)^2\), which agrees quantitatively well with the value of \(t'\) obtained from Wannierization [see \Fig{fig:fig2}(d)]. This hopping renormalization also explains the absence of the zero-energy mode in the spectrum shown in \cref{fig:fig2}(b).

In contrast, for \(v/w > 1\), corresponding to the trivial phase of the SSH interconnect, repeating the same Wannierization procedure shows that the metallic chain effectively fractures at the junction [see upper inset of \Fig{fig:fig2}(c)].

\begin{figure}
    \centering
    \includegraphics{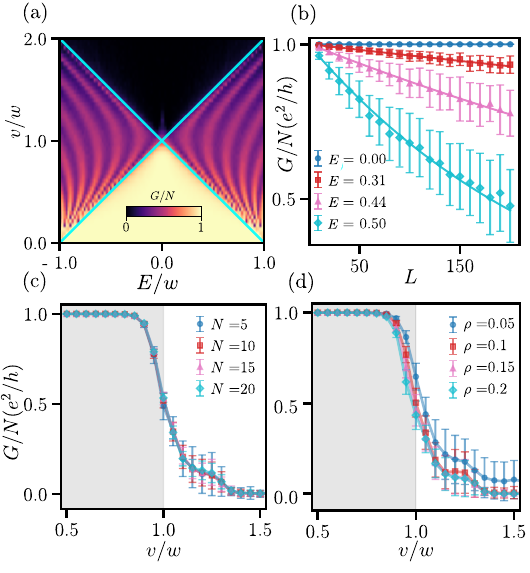}
    \caption{\textbf{Numerical data for multiple wire randomly connected by SSH chains:}
    (a) Phase diagram: Density plot for conductance per wire ($G/N$) as a function of $v/w$ and $E/w$ ($N=5$, $\rho=0.1$ and $L=50$).
    (b) Behaviour of $G/N$ with $L$ for different values of $E$. ($N=5$, $\rho=0.1$, $v/t=0.5$). Solid lines are corresponding best fit curves with the form $Ae^{-L/\xi}$. 
    (c) $G/N$ as a function of $v/w$ for different numbers of wires $N$ for a fixed $\rho=0.1$ and $L=50$.
    (d) $G/N$ as a function of $v/w$ for different $\rho$ ($N=5$, $L=50$).
    For the plots (c) and (d),the shaded region denotes the $v/w$ range where the SSH chains are topological.
     For all of the plots, $w/t=1$, and the $G/N$ is averaged over $100$ configurations. The number of sites in the SSH chain is taken to be $L_{\mathrm{SSH}}=30$.}
    \label{fig:fig3}
\end{figure}

\begin{figure}
    \centering
    \includegraphics{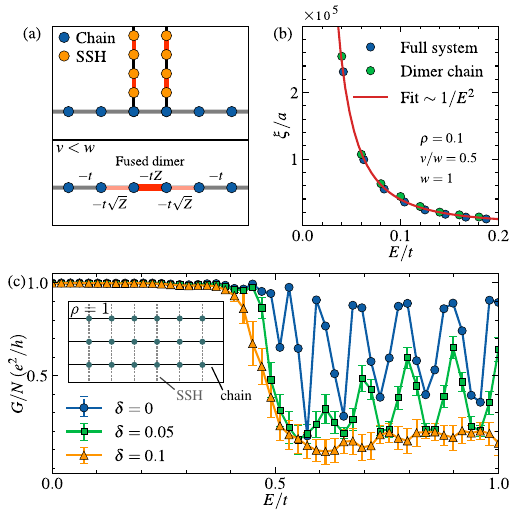}
    \caption{ \textbf{Random-dimer and dense limit:}
(a) 
When two SSH interconnects sit on consecutive sites (top), the effective dimers fuse and renormalize the hoppings of three consecutive bonds (bottom).
This can be generalized for a higher number of fused dimers.
(b) The localization length $\xi$ extracted from \cref{fig:fig3}(b) and from the effective random–dimer hopping model showing the same qualitative divergence $\xi \sim 1/E^2$.
(c) Conductance as a function of energy at $\rho = 1$ (system shown in inset), with $v = \bar{v} + \delta v$ and $w = \bar{w} + \delta w$, where $\delta v$ and $\delta w$, both are randomly sampled from the interval $[-\delta/2, \delta/2]$.
The calculation is done for $N=5$, $L=20$, $L_{\mathrm{SSH}}=30$, $\bar{v}/t=0.5$ and $\bar{w}/t=1.0$. Conductance is averaged over $200$ configurations.}
\label{fig:fig4}
\end{figure}
{\it Network Transport:} 
Next, we discuss the transport properties of the full network shown in \cref{fig:fig1}(a). We compute the two-terminal conductance per wire, \(G/N\), in the \((E/w, v/w)\) plane, as shown in \Fig{fig:fig3}(a).  We consider a relatively high interconnect density \(\rho \approx 0.1\), so that an electron encounters, on average, one SSH interconnect every ten lattice sites of the metallic chain. The four quadrants separated by the bounding lines $E = \pm(v-w)$, exhibit qualitatively distinct transport features.  At $E=0$, when \(v > w\), the conductance vanishes, \(G/N \to 0\). This is the fM phase in which the wires are transport insulators despite a finite density of states at the chemical potential. In contrast, when \(v < w\), we find that \(G/N \to 1\), independent of \(L\) (see \Fig{fig:fig3}(b)), and is referred to as the bM phase. This indicates the absence of back-scattering and perfectly ballistic transport, even in the thermodynamic limit. The perfect transmission $G/N=1$ is remarkable because the system evades Anderson localization even in the presence of randomly placed SSH interconnects in this network and is independent of $N$ and $\rho$ as well [see \Fig{fig:fig3}(c) and (d)]. 

As we move away from the bM phase to $E \neq 0$, we first find the $\mathrm{AI_{ZM}}$ phase. In this phase, the Anderson localization sets in, leading to exponentially decaying conductance \(G/N \sim e^{-L/\xi}\) with system size. The localization length \(\xi\) is, however, very large and diverges as \(\xi \sim 1/E^2\) (see \cref{fig:fig4}(b)). 
At higher energies, the system enters the \(\mathrm{AI_{B}}\) phase, where metallic states hybridize with SSH bulk modes, producing oscillations in the conductance. The positions of these peaks are approximately located at the SSH bulk eigenstates (see \cref{fig:fig2}(b)).

The existence of complete delocalization, free from any scattering at $E=0$, which is independent of interconnect density, is rather unexpected from just symmetry grounds. The system belongs to the BDI symmetry class, which is known to exhibit anomalous localization behavior at the band center \cite{Dyson_1953,Theodorou_1976,Fleishman_1977,Eggarter_1978,Soukoulis_1981,Roman_1987}. While a delocalized state at \(E = 0\) is therefore anticipated, its perfect conductance is not. In generic disordered BDI systems, the conductance at the band center decays with system size as \(G \sim e^{-\sqrt{L/\xi}}\) \cite{Soukoulis_1981}. The absence of such decay in our case indicates that, despite the random placement of interconnects, the effective disorder is not uncorrelated. We discuss the origin of this behavior next.

{\it Mapping to Random Dimer Model:} 
In the trivial phase, the wires are effectively fractured, and the physics is straightforward. We therefore focus on the case where the SSH interconnects are in the topological phase. From the single-junction analysis, an SSH interconnect is known to renormalize two consecutive hopping amplitudes on the wire, \(t \to t\sqrt{Z}\), producing an effective \emph{dimer hopping} at the junction. Consequently, the hopping sequence along the chain becomes \(\ldots, t, t \sqrt Z, t \sqrt Z, t, \ldots\). 
At low interconnect density, these dimers are well-separated and act independently, reducing the system to isolated dimers, corresponding to the well-known \emph{random dimer model} \cite{Dunlap_1990,Wu_1992,Izrailev_1996}. As the density increases, configurations arise in which two interconnects attach to consecutive sites. The corresponding dimers then \emph{fuse}, and the \emph{weight} on their shared bond is multiplied, yielding the hopping sequence \(\ldots, t, t\sqrt Z, t Z, t \sqrt Z, t, \ldots\) [see \cref{fig:fig4}(a)]. Generalization for three or more consecutive interconnects is straightforward.

Although the interconnects are placed randomly, the resulting effective hoppings along the wires are not fully random but instead form short-range correlated off-diagonal disorder. For such a disorder, the wavefunction at \(E = 0\) has a diverging localization length, resulting in perfect transmission (\(G/N = 1\)). Furthermore, in the vicinity of \(E = 0\), the localization length follows $\xi \sim 1/E^2$ \cite{Dunlap_1990,Izrailev_1996}, and there are \(\sim \sqrt{L}\) number of states with localization length exceeding the length of the chain (\(\xi > L\)), which also leads to \(G/N \approx 1\).
Numerically, we compute the localization length both from the exact network model and from an effective random dimer hopping model [see \cref{fig:fig4}(b)]. The two results agree very well, confirming that the system indeed behaves as a network with correlated disorder.

{\it Dense limit:} 
We next examine transport in the perfect density limit where each site of metallic wires is coupled to the adjacent metallic wire by parallel SSH interconnects [see inset in \Fig{fig:fig4}(c)]. We find that $G/N$ remains quantized at unity in this limit as well. This is clearly a {\it weak metal} which conducts in the longitudinal direction but is a transport insulator in the transverse direction. Moreover the conductance is robust to disorder in the $v$ and $w$ as shown in \Fig{fig:fig4}(c)) where we draw both $v$ and $w$ from a uniform distribution between $[\bar{v}-\delta/2, \bar{v}+\delta/2]$ and $[\bar{w}-\delta/2, \bar{w}+\delta/2]$ respectively. The conductance is pinned to unity until the $E \lesssim \abs{\bar v - \bar w}$. Beyond that, we observe pronounced oscillations in the conductance that arise due to the admixing between the metal and SSH bulk bands. However, the localization properties depend sensitively on the $\delta$. Generically, for larger values of $\delta$ the states localize even more.

{\it Outlook:}  Understanding of novel phases of matter out of crystalline systems is a dominant theme in modern condensed matter physics, not only due to the conceptual questions it poses, but also its promise in the advent of a new generation of technological devices. Motivated by the nature of motifs in printable conducting inks \cite{Cao_2008}, we have explored the nature of transport in a bundle of metallic wires that are interconnected via topological SSH interconnects. Proposing a minimal model, we show that while the interconnects are in the trivial insulating phase, they effectively fracture the wires, leading to a transport insulator. Interestingly, however, when the interconnects are in the topological phase, they lead to a dissipation-free transport which is robust to both interconnect density and disorder. We showed that this is a direct result of the mapping of the system to effectively a model of short-range correlated disorder of dimer hoppings. While our results have focused on the non-interacting physics, it will be interesting to explore the system in the presence of interactions where each of the wires effectively behaves as a Luttinger liquid \cite{Benouit_PRB_2005}. Interestingly, Hubbard interactions at the junctions of the interconnects and the wires may host some non-trivial Kondo physics, which will be an important future prospect. The realization of these systems, their first-principles modeling, and quantitatively accurate transport calculations are exciting prospects to explore.

{\it Acknowledgements: } 
AA acknowledges funding from projects IITK/PHY/2022010 and IITK/PHY/2022011. 
AA acknowledges the kind hospitality of ICTS, Bangalore via the Associate Programme, where part of this work was done.
NB acknowledges support from the Institute Postdoctoral Fellowship, IIT Kanpur.
RK acknowledges support from the FARE program at IIT Kanpur.


\bibliography{refs}
\end{document}